\documentclass{aa}
\usepackage{txfonts,hyperref,graphicx}
\pdfoutput=1

\begin{document}

\title{Square root two period ratios in Cepheids and RR Lyrae}

\author{Michael Hippke\inst{1}
  \and John Learned\inst{2}
	\and A. Zee\inst{3}}

\institute{Institute for Data Analysis, Luiter Str. 21b, 47506 Neukirchen-Vluyn, Germany \email{hippke@ifda.eu}
  \and High Energy Physics Group, Department of Physics and Astronomy, University of Hawaii, Manoa 327 Watanabe Hall, 2505 Correa Road Honolulu, Hawaii 96822 USA \email{jgl@phys.hawaii.edu}
	\and Kavli Institute for Theoretical Physics, University of California, Santa Barbara, California 93106 U.S.A. \email{zee@kitp.ucsb.edu}} 

\date{Draft 05 Sep 2014}

\abstract{We document the presence of a few Cepheid and RR Lyrae variable stars with previously unrecognized characteristics. These stars exhibit the property of a period ratio of main pulsation divided by secondary pulsation ($P_{1}/P_{2}$) very close to $\sqrt{2}$.  Other stars of these types have period ratios which do not show clustering with a close association and a single remarkable non-harmonic number. Close examination reveals a deviation of multiples of a few times $\approx$0.06\% for these stars. This deviation seems to be present in discrete steps on the order of $\approx0.000390(4)$, indicating the possible presence of a sort of fine structure in this oscillation.}

\keywords{Stars: variables: Cepheids, Stars: variables: RR Lyrae}

\maketitle

\section{Introduction}
Most classical Cepheids and RR Lyrae are single-mode periodic and radial pulsators. For many years, the only exception known from this simple understanding was a long-term lightcurve modulation, discovered by and named after \citet{Blazhko1907}, found among some RR Lyrae. With better quality of photometry, it was found that various modes of multi-mode pulsations exist within these variables. For an introduction to these oscillations, see \citet{Moskalik2014}. In brief, there are several relevant classes:

First, there are F+1O (fundamental and first overtone) double-mode Cepheids \citep{Osterhoff1957a,Osterhoff1957b} with periods ratios $P_{2}/P_{1}$=0.694 -- 0.746. Several hundreds of these are known within our galaxy and the Magellanic Clouds (e.g. \citet{Sos2008b,Soszynski2010a,Soszynski2011b,Soszynski2012,Marquette2009,Smolec2010}), and in M31 \citep{Poleski2013a} and M33 \citep{Beaulieu2006}. 

Second, many Cepheids are known with a period ratio of $P_{2}/P_{1}$=0.80 -- 0.802, known as 1O+2O (first and second overtone) double-mode Cepheids. Since the discovery by \citet{Mantegazza1992}, about 500 such stars have been found in the Magellanic Clouds \citep{Sos2008b,Soszynski2010a,Soszynski2012,Marquette2009} and only 19 within our galaxy \citep{Smolec2010,Soszynski2011b}. This implies a lower occurrence rate in our galaxy, perhaps due to the higher metallicity within our galaxy. 

Third, there are F+1O double-mode RR Lyrae stars (RRd). Since the first discovery by \citet{Jerzykiewicz1977}, ~2,000 RRd stars have been found in the Magellanic Clouds
\citep{Soszynski2009,Soszynski2010b,Soszynski2012} as well as in our galaxy and in many globular clusters  \citep{Wils2010,Poleski2013b}. This class generally clusters around a period ratio $P_{2}/P_{1}$=0.742 -- 0.748, with a few exceptions of stars with high metallicity, and a ratio as low as 0.726 \citep{Soszynski2011a}.

Fourth, a very few examples of triple mode pulsators among Cepheids and RR Lyrae have been found \citep{Sos2008a}. The most frequent sub-class are F+2O RRd Lyrae-type stars. Nine such stars are known, with period ratios $P_{3}/P_{1}$ = 0.582 -- 0.593 (see \citet{Moskalik2013} and references therein).

Fifth, with precise time-series photometry such as OGLE and Kepler, low-amplitude (a few percent) secondary pulsations have been found, commonly believed to originate from non-radial modes. These modulations occur among Cepheids and RR Lyrae in the range $P_{2}/P_{1}$=0.60 -- 0.64 (\citet{Moskalik2013,Moskalik2014} and references therein).

Differences of period ratios within one group are attributed to differences in metallicity. As metals in stars are present in various compositions and amounts, period ratios are not believed to exist in discrete steps, but in continuous distributions \citep{Buchler2007}. In this paper, we argue that there might be one sub-class of F+1O Cepheids and RR Lyrae that show period ratios in small, discrete steps, very close to $P_{2}/P_{1}=1/\sqrt{2}$. One might wonder, and rightly so, how stars should be able to calculate a square root, obey the calculations and adjust their pulsations accordingly. Let us elaborate.

\section{Motivation and method}

\subsection{Motivation}
Simple dynamical non-linear models can reproduce the frequency content and the asymmetry of variable star light curves well \citep{Hippke2014}. These models generally involve irrational numbers, usually the \textit{golden ratio}, the most irrational number (as it has the slowest continued fraction expansion convergence of any irrational number) and $\sqrt{2}$, sometimes coined \textit{Pythagoras' constant}, the first number to be proven irrational. This has inspired us to search for double-mode variable stars with the same period ratio.

\subsection{Search method}
We have used the largest available variable star catalogs, namely OGLE-II and OGLE-III, EROS \citep{Afonso1999}, and MACHO \citep{Alcock1995}. At the time of writing, OGLE-IV and the Catalina survey have not yet published their data. These surveys are publicly available and publish all found and significant period ratios in their catalogs. Figure~\ref{fig:petersen}) shows all OGLE data. From a literature review, we have collected data for multiperiodic stars in M31 and M33. We have also retrieved the timeseries photometry for all candidate stars and recalculated the frequencies. 

\section{Results}

\subsection{Search results}
\begin{figure}
\resizebox{\hsize}{!}
{\includegraphics{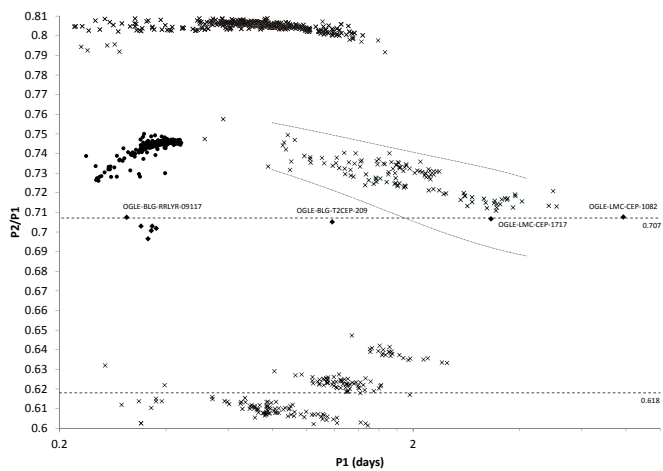}}
\caption{Petersen diagram of main pulsation period versus period ratio for all OGLE-III stars in this range. Crosses and dots represent Cepheids and RR Lyrae, respectively. The skewed lines indicate stability limits for the inclosed Cepheid population \citep{Buchler2007}. The four diamond stars labeled with their catalog designations are closest to the period ratio of $1/\sqrt{2}$ (upper horizontal line). The lower horizontal line represents the golden ratio.}
\label{fig:petersen}
\end{figure}

\begin{figure}
\resizebox{\hsize}{!}
{\includegraphics{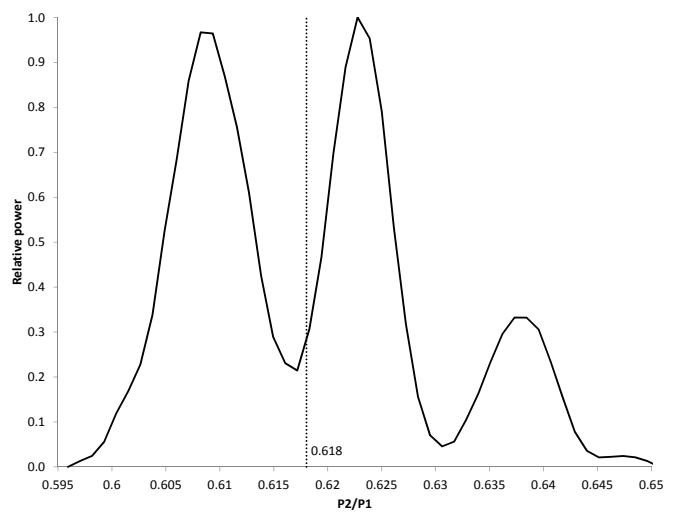}}
\caption{Kernel density estimate for the OGLE stars betwen 0.595 and 0.650 with the golden ratio shown as the dotted vertical line.}
\label{fig:kernelgolden}
\end{figure}

We have found no star with a period ratio equaling the golden ratio (see Figure~\ref{fig:petersen}). This number splits two populations of RR Lyrae stars, as shown in a kernel density estimate (Figure~\ref{fig:kernelgolden}). We have chosen the kernel density estimate \citep{Rosenblatt1956,Parzen1962} instead of a histogram to avoid bin size decisions due to the low number of stars. If real, this split could possibly indicate some sort of watershed in convergence towards and from this irrational period ratio. However, the split is not exact, and it has been argued \citep{Dziembowski2012} that the period ratio clustering among these RR Lyrae is caused by an f-mode instability with high angular degree of \textit{l}=42, 46, 52. The location of these stars above and below the golden ratio could therefore be a simple coincidence, and further modeling will be required to decide this question. We have searched for a fine structure in the stars in this region, but found nothing.

\begin{figure}
\resizebox{\hsize}{!}
{\includegraphics{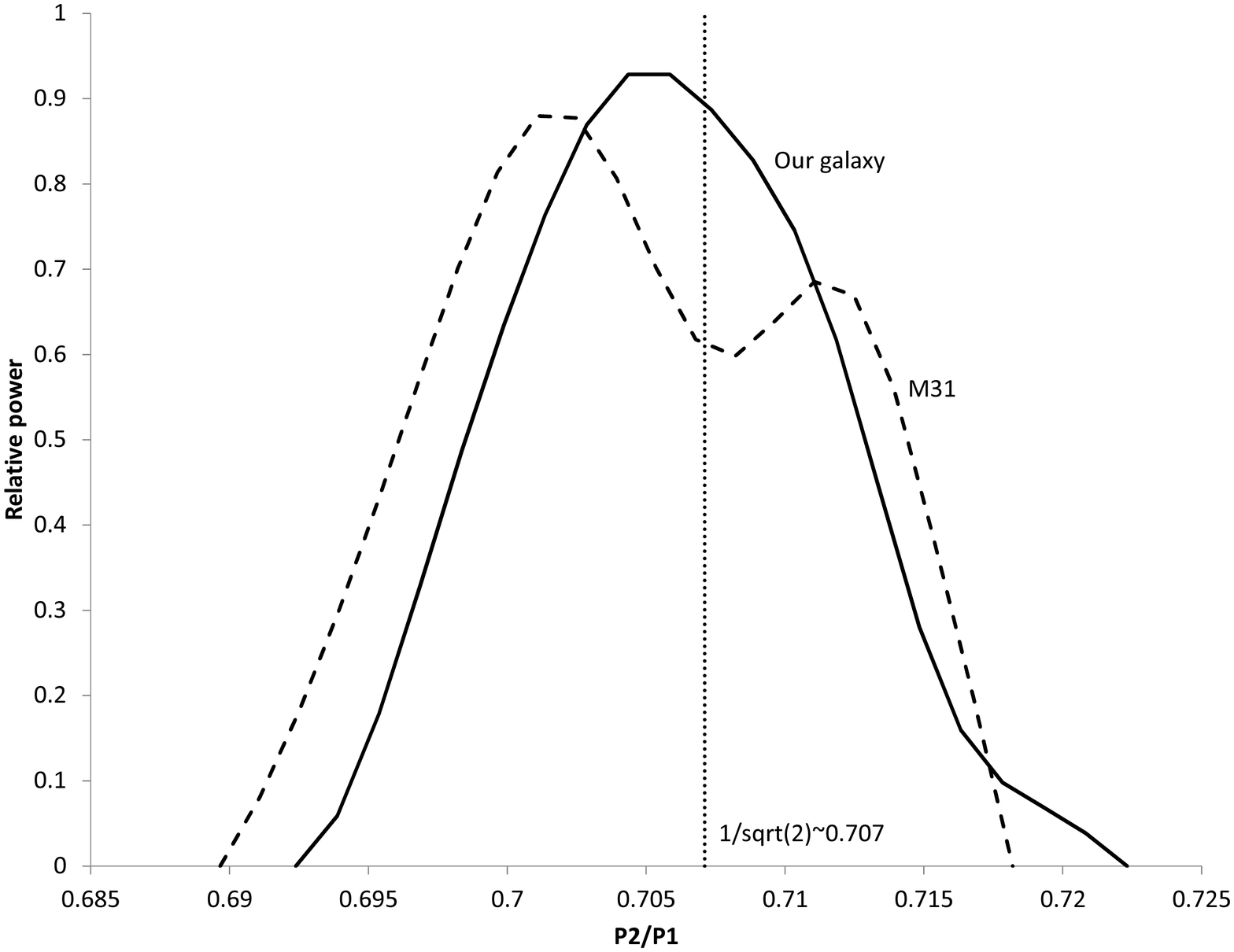}}
\caption{Solid line: kernel density estimate for the 20 stars towards the galactic bulge \citep{Welch2004}. These stars seem to center around $1/\sqrt{2}$, shown by the vertical dotted line. Dashed line: kernel density estimate for the 17 stars in M31 \citep{Lee2013}. These have the same center, but seem to spare the exact value. When using alternative kernels (e.g. gaussian instead of epanechnikov), the effect can also be seen in the galactic bulge stars.}
\label{fig:kernel}
\end{figure}

We then focused our search on $1/\sqrt{2}$, the region of F/1O Cepheids and RR Lyrae. From an extensive literature review, we have only three papers listing stars in the vicinity. The first summarizes the Cepheids in our galaxy (see \cite{Wils2004} and references therein). We have obtained their original raw data \citep{Welch2004}, which is of fair to low quality -- in one case (DZ CMa) only 28 data points were retrieved, which does not allow for a stringent determination of the secondary pulsation. While the uncertainties in the period ratio might be large in these data, one could assume that they cancel out in a distribution graph. In Figure~\ref{fig:kernel}, we have plotted a kernel density estimate for these 20 F/1O Cepheids, and it seems that their center is indeed $1/\sqrt{2}$. This would mean that there is a class of Cepheids (and possibly RR Lyrae) which clusters at this period ratio, and is more abundant in our own galaxy. This is possibly an effect of higher metallicity within our galaxy, as compared to the Magellanic Clouds. 

The second paper presents 5 Cepheids in M33 \citep{Beaulieu2006}, but also suffers from a low number of only 33 measurements. The third paper shows 17 Cepheids in M31 \citep{Lee2013}, and one star with a period ratio of 0.706 caught our attention. Closer examination, however, reveals that this ratio could be as low as 0.699, when using the additional I-band photometry, bringing it further away from $1/\sqrt{2}$. We have thus excluded these data from Figure~\ref{fig:petersen}, but also show the kernel density in Figure~\ref{fig:kernel}. It seems that these also cluster around $1/\sqrt{2}$, however spare the exact number.

The next step was the analysis of the few hundred Cepheids and RR Lyrae from the OGLE catalogs. At first glance, our search for stars with a period ratio of $\sqrt{2}$ has also delivered no hit, meaning that no star possesses this exact ratio, within its errors. However, we have discovered a few stars that come very close and have analyzed them in more detail. These come from the OGLE catalogs, as MACHO and EROS have no stars in the vicinity with period ratios lower than 0.712.

\subsection{Closest candidates}
\begin{figure}
\resizebox{\hsize}{!}
{\includegraphics{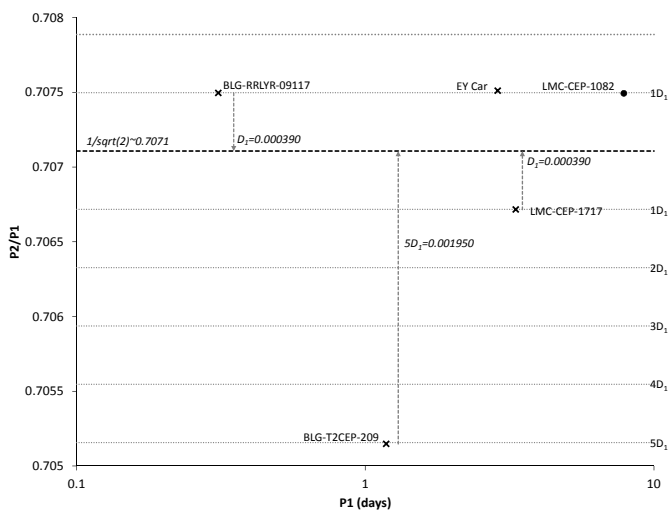}}
\caption{Spacings of the stars with period ratios close to $1/\sqrt{2}$. Symbol sizes approximate uncertainties. Dotted horizontal lines in the upper half of the figure indicate integer multiples of the suggested spacing $D_{1} \approx 0.000390$.}
\label{fig:spacing}
\end{figure}

We have found one RRd Lyrae and four Cepheids (of F/1O type) in the very closest vicinity of the period ratio $P_{2}/P_{1}$ = $1/\sqrt{2}$. Their main pulsation period $P_{1}$ spans a wide range of 0.3 days for the shortest RR Lyrae pulsator and 7.9 days for the longest Cepheid period (see Table~\ref{tab:table1}. When plotting $P_{1}$ versus the period ratio, $P_{2}/P{1}$ (see Figure~\ref{fig:spacing}), there is apparently a fine structure present. We define the deviation from (half) square root two as $D_{X}=(P_{2}/P_{1}) - 1/\sqrt{2}$. The values for $D_{X}$ range from $+0.000391(2)$ (BLG-RRLYR-09117, $0.06\%$) to $-0.001959(5)$ (BLG-T2CEP-209, $0.3\%$). When using LMC-CEP-1717 as a reference ($1D_{1}$), we find BLG-RRLYR-09117 having $-1.004 D_{1}$, LMC-CEP-1082 as $-0.991 D_{1}$, EY Car as $-0.965 D_{1}$ and BLG-T2CEP-209 showing $+5.026 D_{1}$.

While the proximity of the period ratios to $\sqrt{2}$ seems solid, since the errors are small compared to the spacing, we must be clear that the fine structure is at best suggestive at this time. In the following, we will analyze error estimates and potential data glitches.

\begin{table*}
\caption{Candidate stars\label{tab:table1}}
\begin{tabular}{lcccrl}
\hline 
Designation & $P_{1}$ (days)&$P_{2}$ (days)&$R=P_{2}/P_{1}$&$D_{x}=R-1/\sqrt{2}$&Remark \\
\hline 
LMC-CEP-1717\tablefootmark{a}     & 3.32153  & 2.34738 & 0.7067169 & $-0.000389(3)$ & reference = $1D_{1}$\\ 		
BLG-RRLYR-09117\tablefootmark{a}  & 0.30968  & 0.21910 & 0.7074980 & $+0.000391(2)$ & $-1.004 D_{1}$\\         
LMC-CEP-1082\tablefootmark{a}     & 7.86600  & 5.56514 & 0.7074930 & $+0.000386$ & $-0.991 D_{1}$, see text\\ 
BLG-T2CEP-209\tablefootmark{b}    & 1.18128  & 0.83298 & 0.7051480 & $-0.001959(5)$ & $+5.026 D_{1}$\\  
EY Car\tablefootmark{c}						& 2.87602  & 2.03481 & 0.7075108 & $+0.000404(8)$ & $-0.965 D_{1}$\\
\hline 
\end{tabular}

\tablefoottext{a}{OGLE II+III+IV data} 
\tablefoottext{b}{OGLE III data} 
\tablefoottext{c}{5 sources, see text}
\end{table*}

\section{Discussion}

\subsection{Data quality}
\begin{figure*}
\centering
\includegraphics[width=0.49\textwidth]{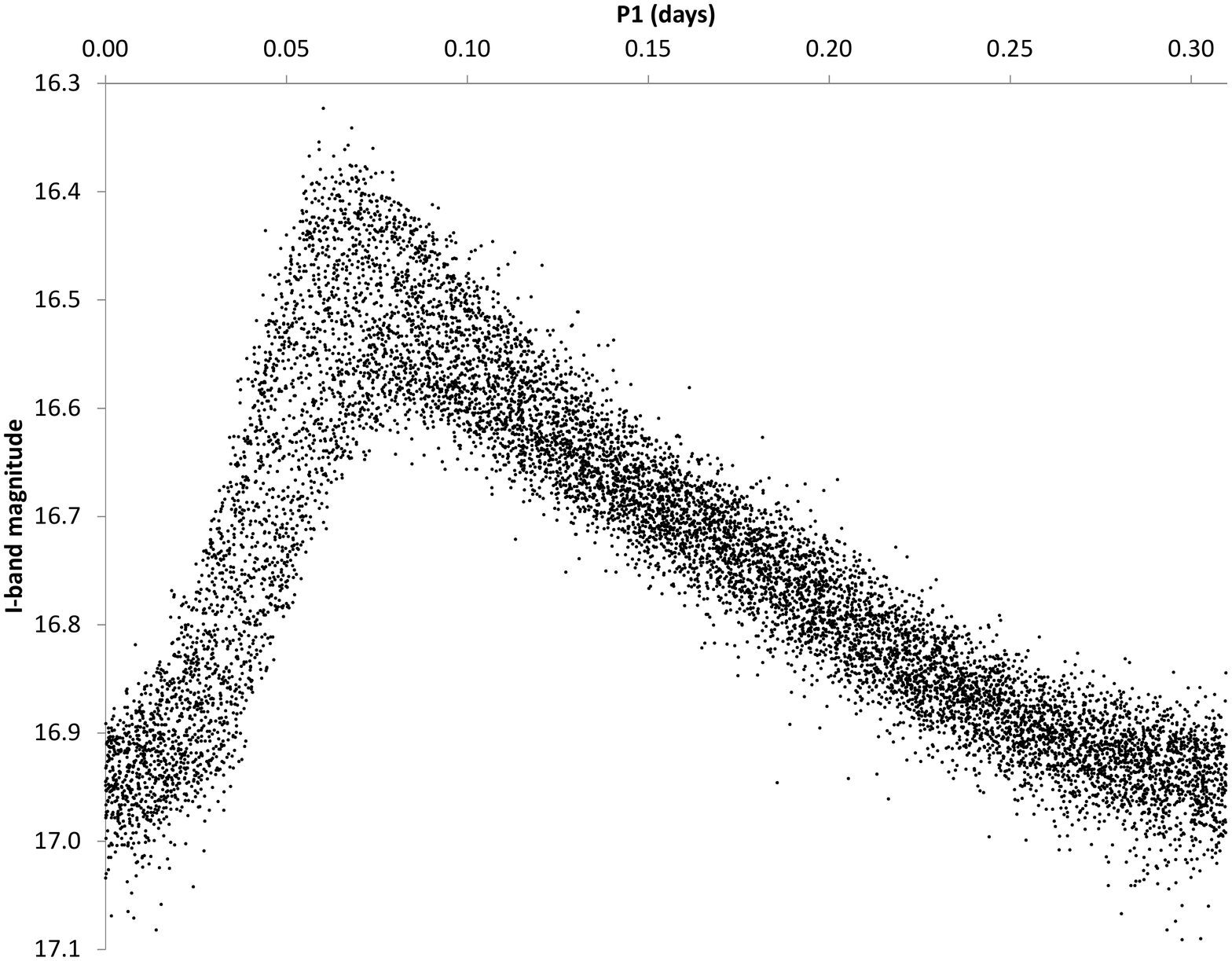}
\includegraphics[width=0.49\textwidth]{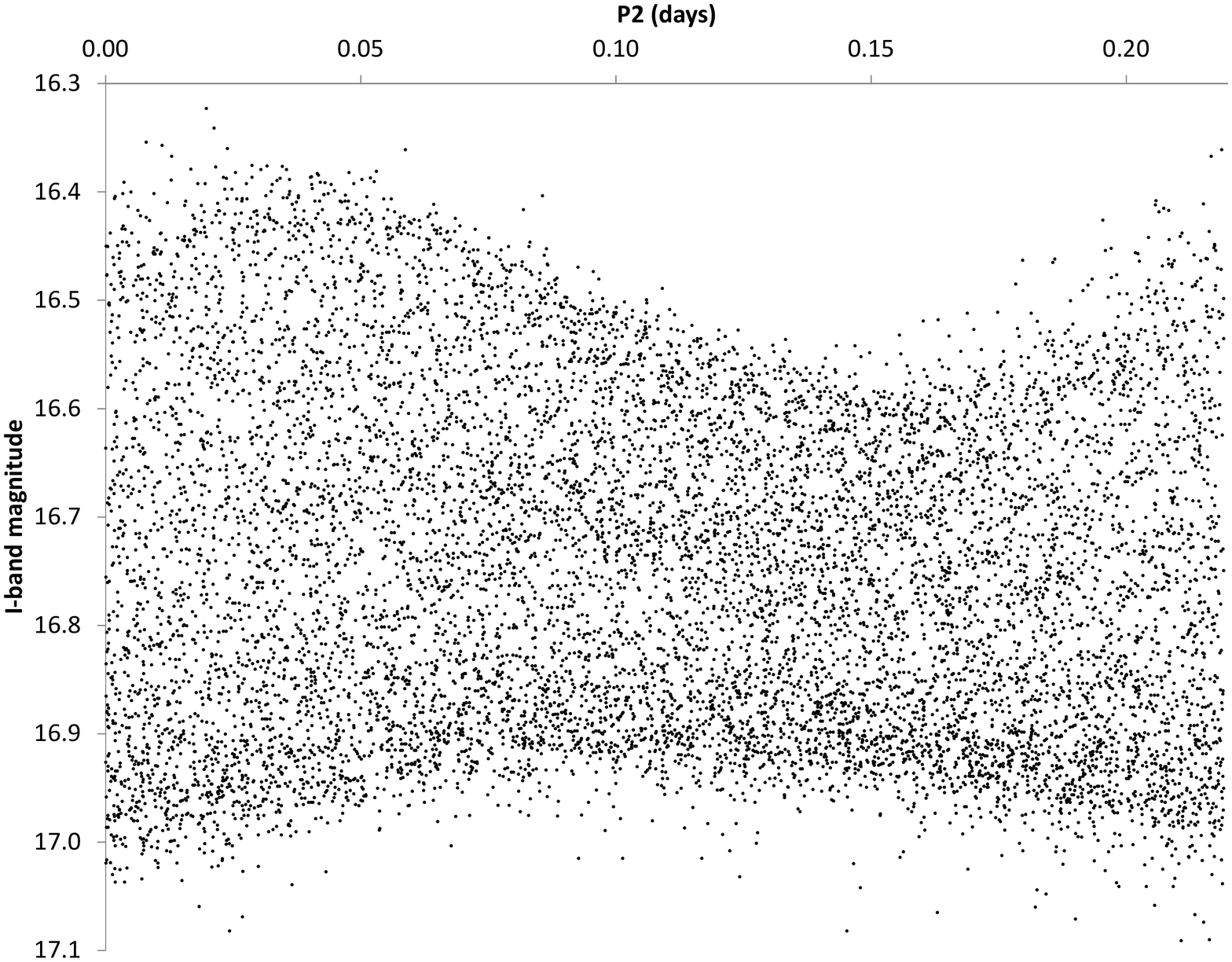}
\caption{Phase fold for BLG-RRLYR-09117 to $P_{1}$=0.30968d (left) and $P_{1}$=0.21910d (right). The amplitude for $P_{1}$ is $\approx$  0.45mag, for $P_{2}$ lower, $\approx 0.2$ mag. Error bars are not plotted, as the spread of points gives a much clearer indication of the true error of the 10,305 data points.\label{fig:fold}}
\end{figure*}

\begin{figure*}
\centering
\includegraphics[width=0.49\textwidth]{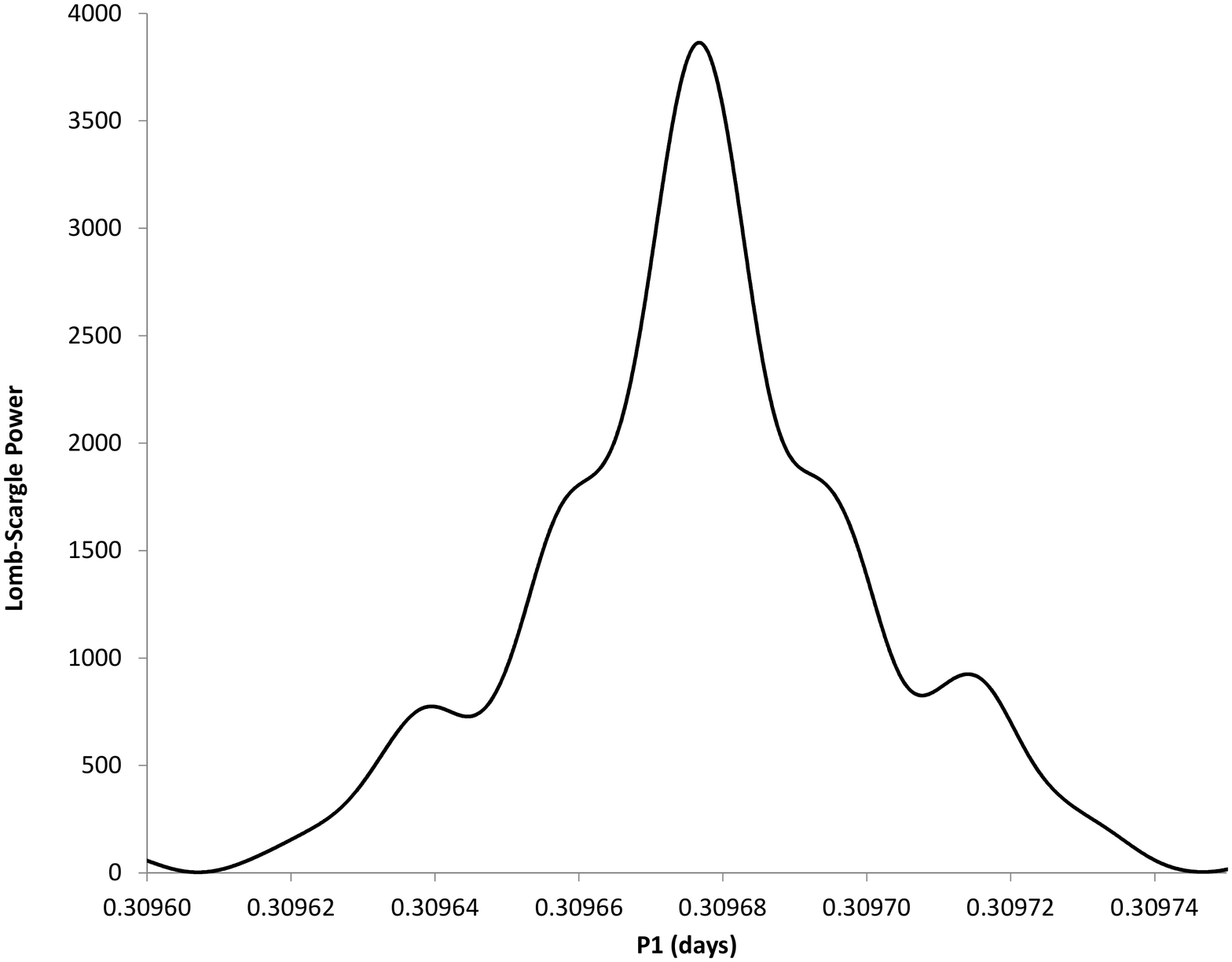}
\includegraphics[width=0.49\textwidth]{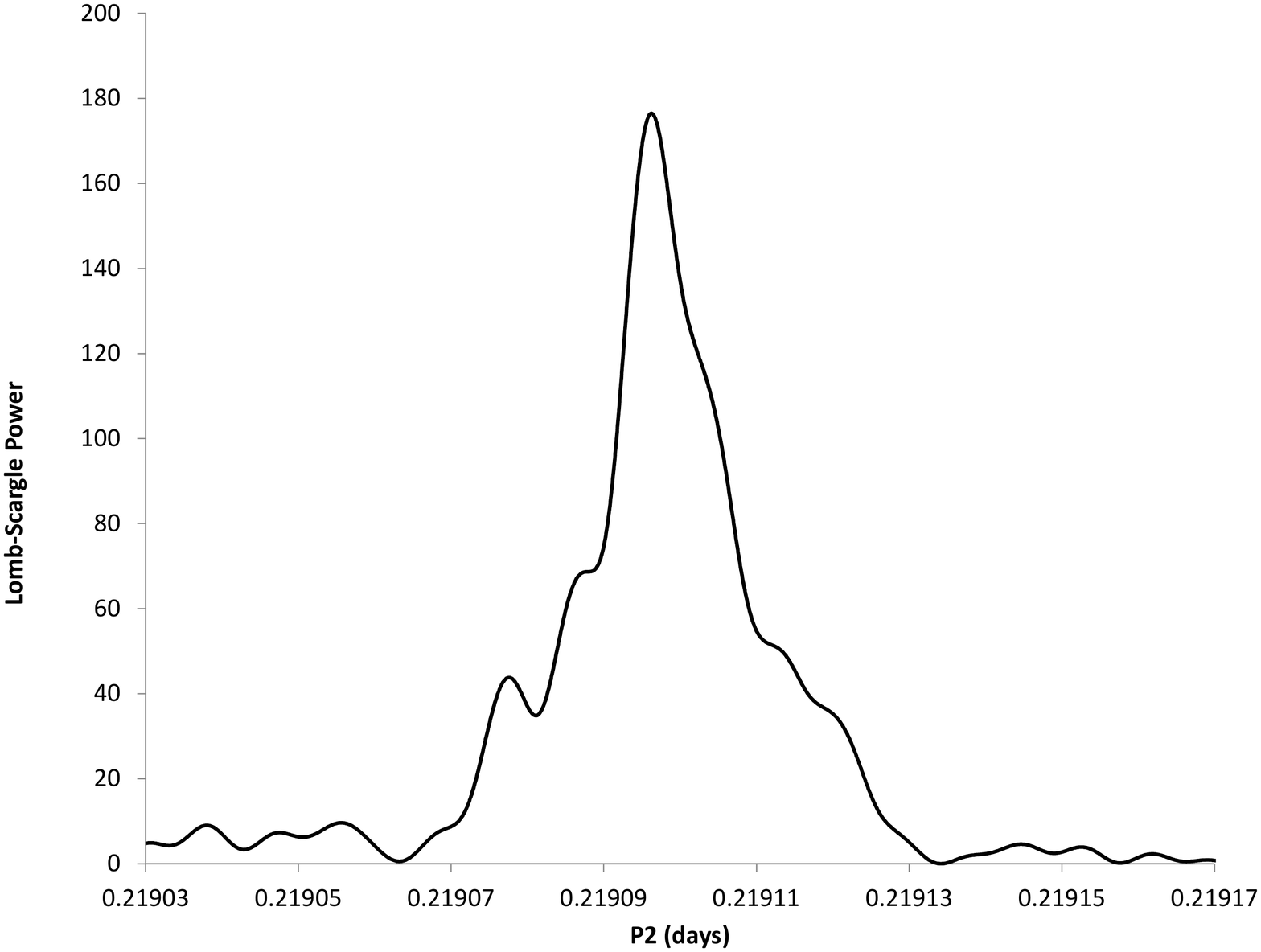}
\caption{Lomb-Scargle periodograms for BLG-RRLYR-09117 focusing on $P_{1}$ (left) and $P_{2}$ (right, without pre-whitening). After removing the primary pulsation, $P_{2}$ is the most significant period. Note different vertical axis.\label{fig:09117kernel}}
\end{figure*}

As with any data, the available photometry has limited precision, which results in statistical uncertainties. To get an impression of the data, we refer to Figure~\ref{fig:fold} for an exemplary phase fold of BLG-RRLYR-09117. Figure~\ref{fig:09117kernel} shows the accompanying Lomb-Scargle periodigrams for $P_{1}$ and $P_{2}$.

For the frequency analysis using Fourier Transforms, \textsc{Period04} \citep{Breger2004} offers a calculation based on the error matrix of a least-squares calculation, as well as a Monte Carlo Simulation. Using LMC-CEP-1082 as an example, both methods typically estimate uncertainties for $P_{1}$ at $6\times10^{-7}$ and for $P_{2}$ at $6\times10^{-6}$. When performing phase folds with such shifted frequencies, one can already visually see the breakdown of the phase fold as increased scatter. This means the estimated uncertainties are probably on the right order.

There is another potential way to derive uncertainties: directly from the period ratios. As described, we have several stars whose deviations are almost identical. For the closest match, BLG-RRLYR-09117 with the reference LMC-CEP-1717, we find $-1.004 D_{1}$. In case these period ratios were in fact identical, the measurement errors would be \textit{a posteriori} on the order of $1\times10^{-7}$. 

Of course this assumes that the fine structure hypothesized is real. 

\subsection{Potential errors}
The easiest (erroneous) explanation for discrete spacings could be the unavoidable quantization in spectral analysis algorithms. There are two main causes for this. The first is based on the simple fact that the algorithms use data types and -structures with a finite bit size, e.g. floating-points in 64-bits. The second comes from limited computing resources, requiring the use of a finite step size in a frequency search. We have analyzed both of these issues in detail.

For the data structures, it can easily be seen that time (e.g. MJD=1000.12345) and flux data (e.g. 10.12345 mag) fit into 64-bits floating points with ease. However, one does not always know which data types the algorithm of the chosen application uses. Also, the required calculations include mathematical functions (mainly sines and cosines) which the application might use via an external system call, and might be rounded there, introducing external error sources. To resolve the data structure issue, we have derived the OGLE photometry as raw data, and merged and reprocessed these using different applications and different methods (\textsc{PERIOD04}, NASA Periodogram Service, and the commercial application \textit{Origin}). We have also calculated all periods separately for OGLE II and OGLE III and compared the values to the OGLE II and III catalog references and found all values to be identical. For the merged data, our results are in between the separate OGLE II and OGLE III values for each dataset. This comes from the additional datapoints and longer time baseline which increase the precision of the measurements. All the results from different algorithms are identical within their errors (with the exception of LMC-Cep-1082, as will be discussed below).

One could still assume a potential data type issue being present in all applications we used. To check this, we have re-written a Lomb-Scargle algorithm in the completely open-source language Free Pascal, using the Lazarus environment. This has the advantage that the complete math unit including sines and cosines can be debugged and modified. By using smaller data types in the subunits (or overloading them with smaller data types), one can even introduce artificial rounding artefacts. With adequate use, we get the exact same results as described above, concluding that today's spectral analysis software does not produce discrete artefacts in period searches.

The second easy cause for such discretization errors could originate from the step size in a frequency search. We chose the same approach as above, with the advantage that the step rate can readily be adjusted in all applications, and the effect can immediately be seen. Using Lomb-Scargle, one can oversample indefinitely without penalty in precision, only at the expense of computing time. Using adequate step rates (e.g. $1\times10^{-8}$), we find that this cannot be causing discrete steps. 

Unfortunately, it is impossible to prove that no other errors exist, and one might argue that the data itself contains a glitch. As all OGLE data are processed the same way, it might be that periodic read-out times of the CCD, or other observational requirements introduce spurious frequencies. To exclude this, one needs data from a completely different source. This has proven difficult, but with an extensive literature search, we have found sufficient data for just one other star in the same period ratio range, EY Car. This Cepheid has been monitored over six decades. We found data from five sources, and treated these equally. We have merged the data, removed obvious outliers (caused by e.g. cosmic ray hits) and subtracted the average zero point. \citet{Mitchell1964} published 23 photoelectric observations using UBV-filters, taken in 1953. We have rejected 3 of these as outliers and thus kept 20. Later, \citet{Pike1979} published 52 observations (2 rejected) taken in 1978. From \citet{Mantegazza1992} we gathered 30 good CCD observations, and 111 observations (1 rejected) from \citet{Berdnikov1995}, taken between 1980 and 1995. Finally, we found 718 (44 rejected) data points from the ASAS-3 survey \citep{Pojmanski2005} between 2003 and 2009. Our period search then included the same checks explained above. The result with Fourier and Lomb-Scargle is $D_{X}$ = +0.000404(8), that is $-0.965 D_{1}$. EY Car has slightly larger estimated errors, but seems to show the discrete step as well.

The sample is still small, and there might be data for one or more other stars in the archives. We encourage any search and analysis.

\subsection{Stability of pulsations and ratios over time}
It is well known that Cepheids and RR Lyrae show period variations over individual cycles \citep{Hippke2014}, or over longer times \citep{Pietrukowicz2001,Poleski2008,Jurcsik2012}. One might therefore argue that the period ratios are a transient phenomenon, vanishing over time. Indeed, all pulsations end one (far) day. With respect to EY Car, we would argue that this generally happens on time scales larger than a few decades.
We have checked the stability of the OGLE data as well, and find them stable over these 18 years (1996 to 2014). However, we had some trouble when calculating the values for LMC-CEP-1082. This ``new FU/FO double-mode pulsator (...) has the longest periods among currently known variables of this class'' \citep{Moskalik2008}\footnote{This reference links to the arXiv preprint of the paper. In the published version, this section was ommited.}, and due to the positive period-luminance relation among Cepheids, this star is particularly bright.

When calculating the secondary pulsation period and the ratio for the OGLE II, III and IV\footnote{After completing the first draft of this paper, we have received the pre-release OGLE-IV data and recalculated all frequencies. The results remained unchanged, the uncertainties decreased. We thank Igor Soszynski for his help.} datasets separately, we got results that were significantly different. We considered three possible causes for this: (1) systematic data errors, (2) deficiencies in the Fourier Transform (FT), and (3) a frequency shift over time. 

As a first step in our analysis, we have calculated Lomb-Scargle periodograms for all datasets and periods, as well as a moving window-function of various lengths, to determine possible time shifts. Our analysis shows that there is considerable jitter over time in both periods and the resulting ratio. It is however unclear whether this stems from instrumental errors or from a physical cause. We also note that LS and FT give significantly different results for the secondary pulsation. The result is also very sensitive to the exact pre-whitening frequency. The derived period ratio deviations are 0.000386 (LS) and 0.000566 (FT). We note that the LS result ($-0.991 D_{1}$) is very close to the proposed discrete spacing, and is considered to be the better method for data with large gaps.

\subsection{Occurrence rate}
With only one RRd Lyrae star (among $\approx$2,000 known) showing a period ratio very close to $1/\sqrt{2}$, the occurrence rate is probably smaller than 1:1,000. Within the F/1O Cepheid group, we found 4 candidates  (among $\approx$500 known), so that the occurrence rate is about 1:100. This estimate suffers from the small sample size, but will be enlarged in the near fututure with new data from time series surveys (e.g. ASAS, OGLE, Catalina)

\subsection{Just a coincidence?}
We have asked the obvious question: how large is the probability that the postulated fine structure is just a coincidence? The sample of stars within $5D_{x}\approx 0.02$ to $1/\sqrt{2}$ consists of only 5 stars, and all of these seem to show the spacing. As the candidate stars are the only ones so close to $P_{2}/P_{1}$ = $1/\sqrt{2}$, we cannot use statistical sample distribution tests. However, we can use the deviations from the fine structure together with the number of candidates. For the OGLE stars, we see a deviation to $D_{x}$ within 1\% (see Table~\ref{tab:table1}). The probability that 4 out of 4 stars fall into this bin is then $2\times1\%^{n-1}$, therefore $0.02^{4-1}$, or 1:125,000. When including the lower data quality star EY Car, with an error of $\approx 3.5\%$ to $D_{x}$, we find the probability as $0.07^{5-1}$, that is 1:41,649.

Admittedly, there is an infinite number of (less and less relevant) irrational numbers. Therefore, there will also be an infinite number of irrational numbers that show a discrete spacing with $n$ stars.

\subsection{Other factors causing unusual period ratios?}
We have considered other factors causing a close connection to an irrational number, and/or the fine structure. In the range $P_{2}/P_{1}$ between 0.65 and 0.71, only one star was found to be discussed in detail in the literature. In a recent paper, the RRab Lyrae \textit{V1127 Aquilae} was studied with data from the CoRoT space telescope \citep{Chadid2010}. A second significant period ($P_{2}/P_{1}=0.6965$) was found and several explanations for this were discussed. As this ratio is relatively close to the ones we discuss here, we review these possibilities. 

One explanation would be a simple blend of \textit{two} stars, either as binaries or just by optical proximity. In dense regions such as the Magellanic Clouds or the Galactic Bulge, this seems not unlikely. However, we argue against this being present in these few stars for two reasons: First, linear combinations of the two frequencies have been found in all candidate stars. Second, it is improbable that all such blends would yield a certain period ratio. The pulsation periods in such variable stars are distributed over a wide range (0.2 days to over 10 days for RR Lyrae and Cepheids, and longer or shorter for other types of variables), which would result in various period ratios not clustering around 0.707. The same is true for the possibility of tidal effects in binary systems, with five of our selected having $P_{1} < 0.3$d. The rotation period seems to be equally unsuited for an explanation. The authors of the above mentioned paper argue that from measured line-widths \citep{Peterson1996}, an upper limit of the rotational velocity of 10-100d can be obtained. The stars discussed herein have shorter $P_{1}$ and $P_{2}$ than 10d.

The remaining explanations are double mode behavior or multi-periodicity, generally assumed to be due to radial and non-radial modes.

\subsection{Linear combinations of frequencies}
When performing a Fourier analysis of these double-mode pulsators, linear combinations of the frequencies are always detected. The strongest are: 
\begin{eqnarray}
f_{3}=f_{1}+f_{2} \label{1} \\
f_{4}=2f_{1}+f_{2} \label{2}
\end{eqnarray}
In the following, we shall neglect all other possible, and usually weaker combinations. It is interesting to note that a frequency ratio of $P_{2}/P_{1} = 1/\sqrt{2}$ (or $f_{2}/f_{1} = \sqrt{2}$) requires these linear combinations to possess the following connection:

\begin{equation}
\frac{f_{4}}{f_{2}} = \frac{f_{3}}{f_{1}} \label{3} \\ 
\end{equation}

This can be shown by inserting (\ref{1}) and (\ref{2}) into (\ref{3}) which yields

\begin{equation}
\frac{2f_{1}+f_{2}}{f_{1}+f_{2}} = \frac{f_{2}}{f_{1}} \\ \label{4}
\end{equation}

which has the trivial solution:

\begin{equation}
\frac{f_{2}}{f_{1}} = \sqrt{2}
\end{equation}

This finding offers no explanation, but a purely phenomenological suggestion on how the $\sqrt{2}$ ratio arises. This finding might allow the improvement of existing linear pulsation models by adding an additional restriction.

\section{Conclusion}
We have found a small set of four Cepheids and one RR Lyrae with period ratios $P_{2}/P_{1} \approx 1/\sqrt{2}$. It is remarkable that these cover the whole range of fundamental periods from short (BLG-RRLYR-09117, $P_{1}=0.31$ days) to the longest known double-mode pulsator (LMC-CEP-1082, $P_{1}=7.87$ days), despite the strong negative connection of period ratios to main pulsation period. It is also remarkable that this is found both among RR Lyrae and Cepheids. We attribute this to the non-linear dynamics in these stars, and look forward to further modeling in this area. 

We found the slight deviations from this special number to be present in integer multiples of the discrete real number $D_{1} \approx 0.000390(2)$. Common pulsation theory has no basis for the occurrence of such a fine structure. It would therefore be convenient to regard these findings as erroneous, or ``in contemporary vernacular, the astronomical community never lets facts stand in the way of a good idea'' \citep{Preston2014}. We believe that our treatment and findings are sound, and encourage the collection and analysis of further data to shed more light on this interesting phenomenon.

\begin{acknowledgements}
AZ's work is supported by the National Science Foundation under grant no. PHY07-57035 and partially by the Humboldt Foundation.
\end{acknowledgements}

\end{document}